\documentstyle [11pt] {article}

\makeatletter

\newcommand{\singlespacing}{\let\CS=\@currsize\renewcommand{\baselinestretch}{1}\tiny\CS}
\newcommand{\oneandahalfspacing}{\let\CS=\@currsize\renewcommand{\baselinestretch}{1.25}\tiny\CS}
\newcommand{\doublespacing}{\let\CS=\@currsize\renewcommand{\baselinestretch}{1.35}\tiny\CS}

\def\@citex[#1]#2{\if@false\immediate\write\@auxout{\string\citation{#2}}\fi
  \def\@citea{}\@cite{\@for\@citeb:=#2\do
    {\@citea\def\@citea{,\linebreak[0]\hskip0pt plus .2em}%
      \@ifundefined{b@\@citeb}%
      {{\bf ?}\@warning{Citation `\@citeb' on page \thepage\space undefined}}%
      \hbox{\csname b@\@citeb\endcsname}}}{#1}}

\newtheorem{rule-def}[theorem]{Rule}

\begin{document}
\newcommand{\la}{\lambda}
\newcommand{\si}{\sigma}
\newcommand{\ol}{1-\lambda}
\newcommand{\be}{\begin{equation}}
\newcommand{\ee}{\end{equation}}
\newcommand{\bea}{\begin{eqnarray}}
\newcommand{\eea}{\end{eqnarray}}
\newcommand{\nn}{\nonumber}
\newcommand{\lb}{\label}

\begin{center}

{\large {\bf {RELATIVISTIC ELECTROMAGNETIC MASS MODELS WITH COSMOLOGICAL VARIABLE
$ {\Lambda} $ IN SPHERICALLY SYMMETRIC ANISOTROPIC SOURCE}}}

\end{center}

\begin{center}

R. N. TIWARI$^{1}$, SAIBAL RAY$^{2}$ AND SUMANA BHADRA$^{3}$\\
$^{1}${\it Department of Mathematics, Indian Institute of
Technology, Kharagpur 721 302, W. B., India}\\ $^{2}${\it
Department of Physics, Darjeeling Government College, Darjeeling
734 101, W. B., India}\\ $^{3}${ \it Department of Physics,
Midnapur College, Midnapur 721 101, W. B., India}\\

\end{center}

\noindent
A class of exact solutions for the Einstein-Maxwell field equations are
obtained  by assuming the erstwhile cosmological constant  $ \Lambda $
to be a space-variable scalar, viz., $ \Lambda =\Lambda(r) $. The source
considered here is static, spherically symmetric and anisotropic charged
fluid. The solutions obtained are matched continuously to the exterior
Reissner-Nordstr\"{o}m solution and each of the four solutions represents
an electromagnetic mass model. \\

\begin{center}
1.INTRODUCTION\\
\end{center}

\noindent
 A very important problem in cosmology is that of the cosmological constant the
present value of which is infinitesimally small $ ( \Lambda \le10^{-56} cm^{-2})$.
However, it is believed that the smallness of the value of $ \Lambda $ at the
present epoch  is because of the Universe being so very old (Beesham$^{2}$).
This suggests that the $\Lambda$ can not be a constant. It will rather be a variable,
dependent on coordinates - either on space or on time or on both (Sakharov$^{30}$;
Gunn and Tinslay$^{15}$;  Lau$^{19}$; Bertolami$^{6}$; Ozer and Taha$^{22}$;
Reuter and Wetterich$^{29}$; Freese et al.$^{12}$; Peebles and Ratra$^{23}$;
Wampler and Burke$^{39}$; Ratra and Peebles$^{26}$; Weinberg$^{40}$; Berman et
al.$^{5}$; Chen and Wu$^{9}$; Berman and Som$^{4}$; Abdel-Rahman$^{1}$;
Berman$^{3}$; Sistero$^{31}$; Kalligas et al.$^{18}$; Carvalho et al.$^{8}$;
Ng$^{21}$; Beesham$^{2}$; Tiwari and Ray$^{37}$). \\
\noindent
 Now, once we assume  $\Lambda$  to be a scalar variable, it acquires altogether a
different status in Einstein's field equations and its influence need not be
limited only to cosmology. The solutions of Einstein's field equations with
variable  $\Lambda$  will have a wider range and the roll of scalar  $\Lambda$
in astrophysical problems will be of as much significance as in cosmology. \\
It is this aspect that motivated us to reexamine the work of Ray and Ray$^{27}$;
and Tiwari and Ray$^{37}$ with the generalization to  anisotropic and charged source
respectively. One can  realize from the present investigations how the variable
$\Lambda$ generates different types of solutions which are physically interesting
as they provide a special class of solutions known as electromagnetic mass
models (EMMM).\\
In section 2,  the Einstein-Maxwell field equations  with variable
$\Lambda$ are provided. Solutions corresponding to different cases for
anisotropic system are given in section 3 . All the solutions obtained are
matched with the exterior Reissner-Nordstr\"{o}m (RN) solution on the boundary of
the charged sphere. Finally, some concluding remarks are made in section 4. \\

\begin{center}
2. FIELD EQUATIONS\\
\end{center}
\noindent
The Einstein-Maxwell field equations for the spherically symmetric metric
\bea
ds^{2}= e^{\nu(r)} dt^{2} - e^{\lambda(r)} dr^{2} - r^{2} ( d \theta ^{2} +
sin^{2} \theta d\phi^{2} )
\eea
corresponding to charged anisotropic fluid distribution are given by
\bea
  e^{-\lambda} ( \lambda^{\prime}/r - 1/r^{2} ) + 1/r^{2} = 8\pi \rho + E^{2}
+ \Lambda,
\eea
\bea e^{-\lambda} ( \nu^{\prime}/r + 1/r^{2} ) -
1/r^{2} = 8\pi {p}_{r} -E^{2} -\Lambda,
\eea
\bea
e^{-\lambda}[\nu^{{\prime}{\prime}}/2 + {\nu^{\prime}}^{2}/4 - {\nu^{\prime}
\lambda^{\prime}}/4 + (\nu^{\prime} - \lambda^{\prime} )/ 2r] =
8\pi p_{\perp} + E^{2} - \Lambda,
\eea
\bea {(r^2 E)}^{\prime} = 4\pi r^2 \sigma e^{\lambda/2}.
\eea The equation (5) can equivalently be expressed in the form,\\
\hspace*{6.0cm} $ E(r) = \frac{1}{r^2} \int_{0}^{r} 4 \pi r^2 \sigma e^{\lambda/2} dr =
\frac{q(r)}{r^2}.$ \hfill $(5a)$\\
\noindent where $q(r)$ is total charge of the sphere under consideration.\\
\noindent
Also, the conservation equation is given by
\bea \frac{d}{dr}(p_r - {\Lambda}/{8 \pi} ) + ( \rho + p_r ) {\nu^\prime}/2 = \frac{1}{8
\pi r^4}\frac{d}{dr}( q^2) + 2( p_{\perp} - p_r)/r.
\eea Here, $\rho, p_r, p_{\perp}, E, \sigma $ and $ q $ are respectively the
matter-energy density, radial and tangential pressures, electric
field strength, electric charge density and electric charge. The
prime denotes derivative with respect to radial coordinate $r$
only.\\
\noindent
Equations. (2) and (3) yield
\bea
e^{-\lambda}(\nu^{\prime} + \lambda^{\prime}) = 8 \pi r( \rho + p_r).
\eea
Again, equation (2) may be expressed in the general form as
\bea
e^{-\lambda} = 1 - 2M(r)/r, \eea where \bea M(r) = 4 \pi
\int_{0}^{r}[ \rho + ( E^2 + \Lambda)/{8 \pi}] r^2 dr
\eea
is the active gravitational mass of a charged spherical body which is
dependent on the cosmological parameter $ \Lambda = \Lambda(r)$.\\

\begin{center}
 3.  SOLUTIONS\\
\end{center}
\noindent A number of solutions can be obtained depending on
different suitable conditions on equation (6). However, as in our
previous work, we assume the relation $ g_{00} g_{11} = -1 $,
between the metric potentials of metric (1), which, by virtue of
equation (7), is equivalent to the equation of state \footnote[2]{
In terms of energy-momentum tensor this can be expressed as $
{T^{0}}_{0} = {T^{1}}_{1} $. } \bea \rho + p_{r}= 0. \eea The
equations (2)-(5) being underdetermined, we further assume the
following conditions \bea \sigma e^{\lambda/2} =\sigma_{0}, \eea
\bea p_{\perp}=n p_{r},\qquad (n\neq 1), \eea where $ \sigma_{0} $
is a constant( which from (5a) can be interpreted as the volume
density of the charge $\sigma$ being constant) and n is the
measure of anisotropy of the fluid system. \\ Equation (5), with
equation (11), then provides the electric field and charge as \bea
E= q/r^{2}= 4 \pi \sigma_{0} r/3. \eea Using equations (10), (12)
and  (13),   in equation (6), we get \bea \frac {d}{dr} (p_{r}-
\Lambda/{8 \pi}) - 2 (n-1) p_{r}/r= 2Ar,\quad A=2 \pi
\sigma_{0}^{2}/3, \eea which is a linear differential equation of
first order. \\ Since the equation (14) involves two dependent
variables, $ p_{r} $ and $ \Lambda $, to solve this equation, we
consider the following four simple cases.\\ \noindent CASE I:\quad
$ \Lambda= \Lambda_{0}-8 \pi p_{r},\quad (\Lambda_{0}= $ constant
) \\ The solutions  in this case are then given by \bea
e^{\nu}=e^{-\lambda}= 1- 2 M(r)/r, \eea \bea \rho= -p_{r}=-
p_{\perp}/n=( \Lambda-\Lambda_{0})/{ 8 \pi}=A a^{-(n-3)} r^{2}
[a^{n-3}-r^{n-3}]/(n-3), \eea \bea M(r) =\frac {4 \pi A a
^{-(n-3)} r^{5}}{15 (n-3)(n+2)} [ (n+2)(n+3)a^{n-3}-30 r^{n-3}]+
\Lambda_{0} r^{3}/6, \eea where $ a $ is the radius of the sphere.
\\ Some general features of these solutions are as follows: \\
\noindent (1) As we want, customarily, $ \rho> 0 $ ( and hence $
p_{r} < 0 $ ), we must have, from (16), $ n >3 $. However, we can
choose $ n < 3 $ ( and certainly $ n \neq 1 $ ). In that case also
$\rho $ becomes positive. This result, viz., the positivity of
matter-energy density is obvious as the electron radius for the
present model is $ ~10^{-13} $ cm, which is much larger than the
experimental upper limit $ 10^{-16} $cm (Quigg$^{25}$). Within
this limit the charge distribution of matter must contain some
negative rest mass (Bonnor and Cooperstock$^{7}$; Herrera and
Varela$^{16}$). This is the reason why we cannot consider $ \rho
\le 0 $ and hence $ p_{r} \ge 0 $ in the present model.\\
\noindent (2) Similarly, we can observe that the effective
gravitational mass (which we get after matching of the interior
solution to the exterior RN solution on the boundary), \bea m =
M(a)+ q^{2}(a)/{2 a} - \Lambda_{0} a^{3}/6 = 8 \pi A (n+3) a^{5}
/{ 5 (n+2)} , \eea is positive for both the choices, $ n > 0 $ and
$ n <0 $. In this respect, the Tolman-Whittaker mass, \bea m_{TW}=
\int_{V} ({T^{0}}_{0}-{T^{\alpha}}_{\alpha}) \sqrt{-g} dV,
(\alpha= 1, 2, 3 \quad  and \quad g \rightarrow 4D)  \nonumber \\
= -\frac {8 \pi A a ^{-(n-3)} r^{5}}{15 (n-3)(n+2)}
[2(n+2)(n+3)a^{n-3}-15(n+1) r^{n-3}] -  \Lambda_{0} r^{3}/3, \eea
can also be examined. In general, this is negative and also equal
to modified Tolman-Whittaker mass (Devitt and Florides$^{10}$),
\bea m_{DF}=e^{-(\nu+\lambda)/2} m_{TW}, \eea as $ \nu+\lambda =0
$, by virtue of the condition $ g_{00}g_{11}=-1 $ in the present
paper. \\ \noindent (3) Pressure being negative the model is under
tension. This repulsive nature of pressure is associated with the
assumption (10), where matter-energy density is positive. This
negativity of the pressure corresponds to a repulsive
gravitational force (Isper and Sikivie$^{17}$; L\'{o}pez$^{20}$).
\\ \noindent (4) The cosmological parameter $\Lambda$, which is
assumed to vary spatially, can be shown to represent a parabola
having the equation of the form $ \Lambda = 8 \pi A [ ( a/2)^2 -
(r - a/2)^2 ] + \Lambda_{0}$ for a particular case $ n =2 $.  The
value of $ \Lambda $ increases from $ 0 $ to $a/2$ and then
decreases  from $a/2$ to $a$ and hence it is maximum at $a/2$ .
The vertex of the parabola is at $ r = a/2$ whereas the values of
$\Lambda$ at $r = 0 $ and at  $r =a$ are $\Lambda_{0}$, the
erstwhile cosmological constant. The same result can also be
obtained from equation (16) as at the boundary of the sphere $r =
a$, $ p_r = p_{\perp}  = 0 $ ( and hence $\Lambda = \Lambda_{0}$
). \\ \noindent (5) The solution set provides electromagnetic mass
model ( EMMM ) (Feynman et al.$^{11}$; Tiwari et al.$^{33, 34,
35}$; Gautreau$^{13}$; Gr{\o}n$^{14}$; Ponce de Leon$^{24}$;
Tiwari and Ray$^{36, 38}$; Ray et al.$^{28}$; Ray and Ray$^{27}$).
This means that the mass of the charged particle such as an
electron originates from the electromagnetic field alone(for a
brief historical background, see Tiwari et al.$^{34}$).\\
\noindent (6) The present model corresponds to Ray and Ray$^{27}$
for $n = 1$, under the assumption $ p_r = - \Lambda/{8 \pi}$. It
can be observed that the other simple possibility, $p_r =
\Lambda/{8 \pi}$, does not exist for this case ( equation (23) of
Ray and Ray$^{27}$. \\ \noindent CASE II:\quad $\Lambda
=\Lambda_{0} + 8 \pi p_r$ \\ \noindent In this case we have the
following set of solutions: \bea e^\nu = e^{-\lambda} = 1-
2M(r)/r, \eea \bea \rho = -p_r = -p_{\perp}/n =
-(\Lambda-\Lambda_{0})/{8 \pi} = Ar^2/(n-1), \eea \bea M(r) = 4
\pi A r^5/15 + \Lambda_{0}r^3/6. \eea \noindent Here some simple
observations are as follows:\\ \noindent (1) In this case also the
electron radius being $ \sim 10^{-13} $ cm the matter-energy
density should be positive (Bonnor and Cooperstock$^{7}$; Herrera
and Varela$^{16}$). This positivity condition requires that $ n $
must be greater than unity.\\ \noindent (2) The effective
gravitational mass , \bea m = 8 \pi A a^5/5, \eea is always
positive whereas the Tolman-Whittaker mass which is also equal to
the modified  Tolman-Whittaker mass , i.e., \bea m_{TW} = m_{DF} =
- 16\pi A r^5/15 - \Lambda_{0} r^3/3, \eea is always negative in
the region $ 0 <r \le a $. The gravitational mass in this case is
independent of anisotropic factor $n$.\\ \noindent (3) The
pressures $ p_r $ and $ p_{\perp} $ are repulsive for $ n>1 $ ( as
in the previous case). \\ \noindent (4) The eq. (22) for $ n=2 $
can be written in the form $ \Lambda = - 8 \pi A r^2 + \Lambda_{0}
$.  This yields a half-parabola whose vertex is at $ r=0 $ and
the parabola lies in the fourth-quadrant of the coordinate systems
$ ( r,\Lambda ) $. \\ \noindent (5) The effective gravitational
mass as obtained in (24) is of
 electromagnetic origin.\\
\noindent
(6) The matter-energy density $ \rho $ as well as the pressures
$ p_r $ and $ p_{\perp} $ are all zero at the centre of the
spherical distribution and increase radially being maximum at
the boundary. This situation is somewhat unphysical though not
at all unavailable in the literature (Som and Bedran$^{32}$).\\
\noindent
CASE III:\quad $ \Lambda = \Lambda_{0} - 8 \pi \int \frac{p_r}{r} dr $ \\
\noindent
The solution set for this case is given by
\bea
e^\nu = e^{-\lambda} = 1- 2 M(r)/r,
\eea
\bea
\rho = - p_r = - p_{\perp}/n
=2Aa^{-(2n-5)}r^2/(2n-5)[ a^{ 2n-5 } - r^{ 2n-5 }],
\eea
\bea
\Lambda =\frac{ 8 \pi A a^{-(2n-5)}r^2}{(2n-3)(2n-5)}[(2n-3)a^{2n-5}-2r^{2n-5}]-
8\pi A a^2/(2n-3)+\Lambda_{0},
\eea
\bea
M(r)=\frac{8\pi A a^{-(2n-5)}r^5}{15n(2n-3)(2n-5)}[n(n+2)(2n-3)a^{2n-5}-15(n-1)r^{2n-5}]  \nonumber \\
-4\pi A a^2r^3/[3(2n-3)]+\Lambda_{0}r^3/6.
\eea
\noindent
Some general features of the above set of solution are as
follows: \\
\noindent
(1) The matter-energy density is positive and pressures are
negative for $ n>5/2 $.\\
\noindent
(2) The effective gravitational mass,
\bea
m = 8 \pi A (3n+1 ) a^{5} /15n,
\eea
is positive for $ n>1 $. On the other hand, the Tolman-Whittaker
mass and the modified Tolman-Whittaker mass, being equal, are given by
 \bea
m_{TW} = m_{DF} =- \frac{32 \pi A
a^{-(2n-5)}r^5}{15n(2n-3)(2n-5)}
[n(n+2)(2n-3)a^{2n-5}-15(n-1)(2n-1)r^{2n-5}  \nonumber \\
+8 \pi A a^{2}r^{3}/[3(2n-3)]-\Lambda_{0}r^3/3.
\eea
\noindent
Depending on the different values of $ n $ these masses may be negative
or positive.\\
\noindent
(3) The matter-energy density and the pressures, as usual, are zero at the
centre  $r=0$ as well as at the boundary $r = a$.  Thus the maximum value
must be in the region $ 0<r<a $. This can be confirmed from equation(27)
which, for the value $ n=2 $, represents a parabola of the form $ \Lambda
= 2A [(a/2)^2-(r-a/2)^2] $, the vertex being at $ r=a/2 $. \\
\noindent
(4) The value of $\Lambda$  at the centre $ r=0 $ is $ [ \Lambda_{0} -8 \pi Aa^2/(
2n-3)]$. It acquire  maximum value $ \Lambda_{0}$  at the boundary $ r=a $.\\
(5) The solution set represents EMMM.\\
\noindent
CASE IV:\quad $ \Lambda = \Lambda_{0} + \int \frac {p_r}{r} dr $ \\
\noindent
The solutions in this case are given by
\bea
e^\nu = e^{-\lambda} = 1-2M(r)/r,
\eea
\bea
\rho = -p_r = -p_{\perp}/n = \frac {2Aa^{-(2n-3)}r^2}{(2n-3)}[a^{2n-3}-
r^{2n-3}],
\eea
\bea
\Lambda = -\frac{ 8 \pi
Aa^{-(2n-3)}r^2}{(2n-1)(2n-3)}[(2n-1)a^{2n-3}-2r^{2n-3}] \nonumber\\
 +8 \pi Aa^2/(2n-1) +\Lambda_{0},
\eea
\bea
M(r) = \frac{8 \pi Aa^{-(2n-3)}r^5}{15(n+1)(2n-1)(2n-3)}[n(n+1)(2n-1)a^{2n-3}
-15(n-1)r^{2n-3}] \nonumber\\
+4 \pi Aa^2r^3/[3(2n-1)] + \Lambda_{0}r^3/6.
\eea
\noindent
Here, the observations are as follows: \\
\noindent
(1) The matter-energy density is positive and pressures are negative for
 $ n>3/2 $. \\
\noindent
(2) The effective gravitational mass,
\bea
m = 8 \pi A(3n+5)a^5/15(n+1),
\eea
for the condition $ n>1 $ is always positive, whereas the Tolman-Whittaker mass,
\bea
m_{TW} =m_{DF} = - \frac{8 \pi Aa^{-(2n-3)}r^5}{15(n+1)(2n-1)(2n-3)}[4n(n+1)(2n-1)a^{2n-3}
-15(n-1)(2n+1)r^{2n-3}], \nonumber \\
 - 8 \pi Aa^2r^3/[3(2n-1)] -\Lambda_{0}r^3/3,
\eea
may have positive or negative value depending on the choice of $ n $. \\
\noindent
(3) The values related to $ \rho $ and $ p $ are zero both at $ r=0 $ and $ r=a
$.\\
\noindent
(4) The effective gravitational mass as well as the other physical variables,
including $ \Lambda$, are of purely electromagnetic origin.\\

\begin{center}
4. CONCLUSION\\
\end{center}
\noindent
(1) As mentioned in the introduction, the present work considers $ \Lambda $,
the erstwhile cosmological constant, to be a variable dependent on space
coordinates. The contribution of this variable $ \Lambda $ can be seen in the
calculations given in the previous sections. It can be seen that $ \Lambda $ is
related to pressure and matter-energy density, and therefore contributes  to
effective gravitational mass of the astrophysical system. \\
\noindent
(2) The present EMMMs have been obtained under the condition $ \rho + p_r = 0 $
(eq.(10)). This problem thus requires further investigation to see whether such
models can be obtained even for the condition $ \rho +  p_r \not= 0 $.\\

\begin{center}
ACKNOWLEDGEMENT\\
\end{center}
SR is thankful to Centre for Theoretical Studies, Indian Institute
of Technology, Kharagpur where a part of this work was carried
out. We all thank referee for his valuable comments that improved
the paper.

\begin{center}
 REFERENCES
\end{center}

\begin{enumerate}
\item{} A.- M. M. Abdel-Rahman, {\it Gen. Rel. Grav.} {\bf 22} (1990), 655.
\item{} A. Beesham, {\it Phys. Rev. D} {\bf 48} (1993), 3539.
\item{} M. S. Berman, {\it Int. J. Theor. Phys.} {\bf 29} (1990a), 567;
 {\it Int. J. Theor. Phys.} {\bf 29} (1990b), 1419;
             {\it Gen. Rel. Grav.} {\bf 23} (1991a), 465; {\it Phys. Rev. D}
             {\bf 43} (1991b), 1075.
\item{} M. S. Berman  and M. M. Som, {\it Int. J. Theor. Phys.}
               {\bf 29} (1990), 1411.
\item{} M. S. Berman, M. M.  Som  and F. M. Gomide,  {\it Gen. Rel.
                Grav.} {\bf 21} (1989), 287.
\item{} O.  Bertolami, {\it Nuo. Cim. B} {\bf 93} (1986a), 36; {\it Fortschr.
Phys.} {\bf 34} (1986b),  829.
\item{} W. B. Bonnor and F. I. Cooperstock, {\it Phys. Lett. A} {\bf 139} (1989), 442.
\item{} J. C.  Carvalho, J. A. S.  Lima  and I. Waga, {\it Phys. Rev.
            D} {\bf 46} (1992), 2404.
\item{} W.  Chen and Y. S. Wu, {\it Phys. Rev. D} {\bf 41} (1990), 695.
\item{} J. Devitt and P. S. Florides, {\it Gen. Rel. Grav.} {\bf 21} (1989), 585.
\item{} R. P. Feynman, R. R. Leighton  and M. Sands, {\it The Feynman
            Lectures on Physics} (Addison-Wesley, Palo Alto, Vol.II,
            Chap. 28) (1964).
\item{} K.  Freese, F. C. Adams, J. A. Frieman  and E. Mottola,
        {\it nucl. Phys. B} {\bf 287} (1987), 797.
\item{} R.  Gautreau, {\it Phys. Rev. D}, {\bf 31} (1985), 1860.
\item{} {\O}.  Gr{\o}n, {\it  Phys. Rev. D} {\bf 31} (1985),  2129;
            {\it Am. J. Phys.} {\bf 54} (1986a),  46; {\it Gen. Rel.
            Grav.} {\bf 18} (1986b), 591.
\item{} J. Gunn  and B. M. Tinslay, {\it Nature} {\bf 257} (1975), 454.
\item{} L. Herrera and V. Varela, {\it Phys. Lett. A}, {\bf 189} (1994),  11.
\item{} J. Isper and P. Sikivie, {\it Phys. Rev. D} {\bf 30} (1983), 712.
\item{} D. Kalligas, P. Wesson and C. W. F. Everitt, {\it Gen. Rel. Grav.}
            {\bf 24} (1992), 351.
\item{} Y. K. Lau, {\it Aust. J. Phys.} {\bf 29} (1985),  339.
\item{} C. A. L\'{o}pez,  {\it Phys. Rev. D} {\bf 38} (1988), 3662.
\item{} Y. J. Ng, {\it Int. J. Theor. Phys.} {\bf 1} (1992), 154.
\item{} M. Ozer and M. O. Taha, {\it Phys. Lett. B} {\bf 171} (1986),  363;
            {\it Nucl. Phys. B} {\bf 287} (1987), 776.
\item{} P. J. E. Peebles and B. Ratra, {\it Astrophys. J.} {\bf 325} (1988),
L17.
\item{} J. Ponce de Leon, {\it J. Math. Phys.} {\bf 28} (1987a),  410;
            {\it Gen. Rel. Grav.} {\bf 19} (1987b), 197;  {\it J. Maths.
            Phys.} {\bf 29} (1988), 197.
\item{} C. Quigg, {\it Gauge theories of the strong, weak and
electromagnetic interactions} (Benjamin, New York) (1983), p. 3.
\item{} B. Ratra and P. J. E. Peebles, {\it Phys. Rev. D} {\bf 37} (1988),
3406.
\item{} S. Ray and D. Ray, {\it Astrophys. Space Sci.} {\bf 203} (1993), 211.
\item{} S. Ray, D.  Ray  and R. N. Tiwari, {\it Astrophys. Space Sci.}
            {\bf 199} (1993),  333.
\item{} M. Reuter and C. Wetterich, {\it Phys. Lett. B} {\bf 188} (1987),38.
\item{} A. D. Sakharov, {\it Doklady Akad. Nauk. SSSR } {\bf 177} (1968),
                70 (translated: {\it Soviet Phys. Doklady} {\bf 12}).
\item{} R. F. Sistero, {\it Gen. Rel. Grav.} {\bf 23} (1991), 1265.
\item{} M. M. Som  and M. L. Bedran, {\it Phys. Rev. D} {\bf 24} (1981), 2561.
\item{} R. N. Tiwari, J. R.  Rao  and R. R. Kanakamedala, {\it Phys.
                Rev.D } {\bf 30} (1984),  489.
\item{} R. N. Tiwari, J. R.  Rao  and R. R. Kanakamedala, {\it Phys.
                Rev. D} {\bf 34} (1986), 1205.
\item{} R. N. Tiwari, J. R.  Rao and S.  Ray, {\it Astrophys. Space
            Sci.} {\bf 178} (1991), 119.
\item{} R. N. Tiwari and S. Ray, {\it Astrophys. Space Sci.} {\bf 180} (1991a),
            143; {\it Astrophys. Space Sci.} {\bf 182} (1991b), 105.
\item{} R. N. Tiwari and S. Ray, {\it Ind. J. Pure Appl. Math.}
        {\bf 27} (1996), 907.
\item{} R. N. Tiwari and S. Ray, {\it Gen.  Rel.  Grav.} {\bf 29} (1997), 683.
\item{} E. J. Wampler and W. L. Burke, {\it New Ideas in
                Astronomy} (Cambridge Univ. Press) (1988), p. 317.
\item{} S. Wienberg, {\it Rev. Mod. Phys.} {\bf 61} (1989), 1.
\end{enumerate}

\end{document}